\documentclass[headings=standardclasses]{scrartcl}
\usepackage[utf8]{inputenc}
\usepackage[T1]{fontenc}

\usepackage[english]{babel}
\usepackage{mathtools,amssymb,amsthm} %math
\usepackage{verbatim,csquotes}
\usepackage{tikz} % Diagrams
\usetikzlibrary{cd} % Commutative Diagrams

\RequirePackage{hyperref}
\hypersetup{breaklinks,unicode=true}

\usepackage{orcidlink}
\usepackage{xcolor}

\theoremstyle{plain}
\newtheorem{theorem}{Theorem}
\newtheorem*{theorem*}{Theorem}

\newtheorem*{lemma*}{Lemma}

\newtheorem*{proposition*}{Proposition}

\newtheorem*{corollary*}{Corollary}
\theoremstyle{definition}

\newtheorem*{definition*}{Definition}

\newtheorem*{example*}{Example}
\theoremstyle{remark}
\newtheorem{remark}{Remark}
\newtheorem*{remark*}{Remark}

\newtheorem*{conjecture*}{Conjecture}

\newtheorem*{problem*}{Problem}

\newcommand*{\dd}{\mathrm{d}}

\DeclareMathOperator{\Id}{Id}
\DeclareMathOperator{\pr}{pr}
\newcommand*{\email}[1]{\href{mailto:#1}{#1}}
\newcommand*{\orcidID}[1]{\orcidlink{#1}}

\title{The Herglotz principle and vakonomic dynamics}
\date{\today}
\begin{document}
\author{Manuel de León\orcidID{0000-0002-8028-2348}
    \thanks{Instituto de Ciencias Matem\'aticas~(CSIC-UAM-UC3M-UCM)
    , Madrid, Spain}
    \thanks{Real Academia de Ciencias Exactas, Físicas y Naturales,  Madrid, Spain} \\\email{mdeleon@icmat.es} \and
     Manuel Lainz\orcidID{0000-0002-2368-5853}\footnotemark[1] \\
     \email{manuel.lainz@icmat.es} \and
    Miguel C. Mu\~noz-Lecanda\orcidID{0000-0002-7037-0248}
    \thanks{Department of Mathematics, Universitat Polit\`ecnica de Catalunya} \\ \email{miguel.carlos.munoz@upc.edu}}
% \authorrunning{M. de Le\'on, M. Lainz and M. C. Muñoz Lecanda}

% \institute{ \and
%  \and
% Department of Mathematics,
% Universitat Polit\`ecnica de Catalunya}

\maketitle

\begin{abstract}
    In this paper we study vakonomic dynamics on contact systems with nonlinear constraints. In order to obtain the dynamics, we consider a space of admisible paths, which are the ones tangent to a given submanifold. Then, we find the critical points of the Herglotz action on this space of paths. This dynamics can be also obtained through an extended Lagrangian, including Lagrange multiplier terms.

    This theory has important applications in optimal control theory for Herglotz control problems, in which the cost function is given implicitly, through an ODE, instead of by a definite integral. Indeed, these control problems can be considered as particular cases of vakonomic contact systems, and we can use the Lagrangian theory of contact systems in order  to understand their symmetries and dynamics.

    \par\vskip\baselineskip\noindent
\end{abstract}
    
\section{Introduction}

Given a Lagrangian $L:TQ \to \mathbb{R}$ and a submanifold $N \subseteq TQ$, one can look for the critical points of the Euler-Lagrange action restricted to the paths which are tangent to $N$. This critical points are the solutions of the Euler-Lagrange equations for the extended Lagrangian~\cite{Arnold1997} $\mathcal{L}(q^i,\dot{q}^i, \lambda_a) = L(q^i, \dot{q}^i) - \lambda_a \phi^a$, where $\{\phi^a \}$ are a set of independent constraints defining $N$. We remark that the dynamics obtained from this principle is, in general, different to nonholonomic dynamics~\cite{cortes2002geometric,gracia2003some}, in which the critical points of the action are computed on the unconstrained space of paths, but the admisible variations are constrained.

While nonholonomic dynamics has applications in engineering problems, vakonomic mechanics can be used to study optimal control problems. This opens up the possibility to apply results and techniques from Lagrangian mechanics to the study of optimal control problems, such as the Noether theorem and its generalizations~\cite{martinez2001symmetries}, or variational integrators constructed from the theory of discrete mechanics~\cite{benito2005discrete}.

On the other hand, in the Herglotz variational principle one considers a Lagrangian $L:TQ \times \mathbb{R} \to \mathbb{R}$, $L(q^i,\dot{q}^i,z)$ that depends not only on the positions and velocities of the system, but also on the action $z$ itself. The action is then defined implicitly, through the ODE $\dot{z}=L(q^i,\dot{q}^i,z)$. The critical points of this action are the solutions of the Herglotz equations~\cite{Georgieva2003,Herglotz1930}:
\begin{equation}
    \frac{\partial L}{\partial q^i} - \frac{\dd }{\dd t} \frac{\partial L}{\partial \dot{q}^i} = \frac{\partial L}{\partial \dot{q}^i}  \frac{\partial L}{\partial z}.
\end{equation}
It has recently been acknowledged that the Herglotz principle provides the dynamics for the Lagrangian counterpart of contact Hamiltonian systems~\cite{deLeon2019}. This has allowed the developement of a theory of symmetries~\cite{deLeon2020b,Gaset2020b} which in this setting are not related to conserved quantities, but to \emph{dissipated} ones, which decay at the same rate as the energy. Furthermore variational integrators based on the Herglotz principle have been developed~\cite{Vermeeren2019,Simoes2021}.

Contact dynamics and the Herglotz principle have applications on the description of many physical systems, such as mechanical systems with friction, thermodynamic systems and some cosmological models~\cite{Bravetti2017,Mrugala1991,sloan2020new,usher2020local}.

While the dynamics of contact systems with (linear) nonholonomic constraints~\cite{deLeon2020d} has been studied, a theory of contact vakonomic dynamics has not still been developed. This theory could be useful for the study of the Herglotz Optimal Control Problem, introduced in~\cite{deLeon2020a}, in which the cost function is defined by an ODE, instead of an integral. This will allow a new way to obtain the dynamical equations (on~\cite{deLeon2020a} they were obtained rather indirectly, throught Pontryaguin maximum principle) and to apply some of the results of contact Lagrangian systems to this situation.

The paper is structured as follows. In Section~2 we review the Herglotz principle. Its traditional formulation, in which the action is defined implicitly, makes the implementation of the constraints difficult. We present and alternative in which the action is defined explicitly, but vakonomic constraints are present. This will make the addition of new constraints almost trivial.
In Section~3 we obtain the vakonomic dynamical equations for a constrained contact system, and see that this dynamics can also be obtained through an extended contact Lagrangian.
Finally, in Section~4, we sketch the relationship between the vakonomic dynamics of contact Lagrangian systems and the Herglotz optimal control problem.

\subsection{The Herglotz variational principle, revisited}
Let $Q$ be the configuration manifold and let $L:TQ \times \mathbb{R} \to \mathbb{R}$ be the contact Lagrangian.

Consider the (infinite dimensional) manifold  $\Omega(q_0, q_1)$  of curves $c:[0,1]\to Q$ with endpoints $q_0,q_1 \in Q$. That is, $c(0)=q_0$, $c(1)=q_1$. The tangent space  of $\Omega(q_0,q_1)$ at the curve $c$, is the space of vector fields along $c$ vanishing at the endpoints. That is,
\begin{equation}
    \begin{split}
        T_c \Omega(q_0,q_1) =  \{
            \delta c:[0,1] \to TQ  \mid &  \delta   c(t) \in  T_{c(t)} Q  \, \text{for all t} \in [0,1], \\ &   \delta c(0)=0, \, \delta c(1)=0  \}.
    \end{split}
\end{equation}

We fix a real number $z_0 \in \mathbb{R}$ and consider the following operator:
\begin{equation}\label{eq:Z_operator}
    \mathcal{Z}:\Omega(q_0,q_1) \to \mathcal{C}^\infty ([0,1] \to \mathbb{R}),
\end{equation}
 which assigns to each curve $c$ the function $\mathcal{Z}(c)$ that solves the following ODE:
\begin{equation}\label{contact_var_ode}
\begin{dcases}
    \frac{\dd\mathcal{Z}(c)}{\dd t} &= L(c, \dot{c}, \mathcal{Z}(c)),\\
    \mathcal{Z}(c)(0) &= z_0,
    \end{dcases}
\end{equation}
that is, it assigns to each curve on the base space, its action as a function of time.

Now, the \emph{contact action functional} maps each curve $c\in \Omega(q_0, q_1)$ to the increment of the solution of the ODE:
\begin{equation}\label{eq:contact_action}
    \begin{aligned}
        \mathcal{A}: \Omega(q_0,q_1) &\to \mathbb{R},\\
        c &\mapsto \mathcal{Z}(c)(1) - \mathcal{Z}(c)(0).
    \end{aligned}
\end{equation}
Note that, by the fundamental theorem of calculus,
\begin{equation}
    \mathcal{A}(c) = \int_0^1 L(c(t),\dot{c}(t), \mathcal{Z}(c)(t)) \dd t.
\end{equation}
Thus, in the case that $L$ does not depend on $z$, this coincides with the classical Euler-Lagrange action.

\begin{remark}
    The Herglotz action is usually defined as  ${\mathcal{A}}_0(c) =  \mathcal{Z}(c)(1)$.
    However, this definition and our definition only differ by a constant.
    Indeed,
    \begin{equation}
        \mathcal{A}(c) = {\mathcal{A}_0}(c) - z_0.  
    \end{equation}
    In particular they have the same critical points. However the computations in the vakonomic principle are simpler for $\mathcal{A}$.
\end{remark}

As it is proved in~\cite{deLeon2019}, the critical points of this action functional are precisely the solutions to Herglotz equation:
\begin{theorem}[Herglotz variational principle]\label{thm:Herglotz_principle}
    Let $L: TQ  \times \mathbb{R} \to \mathbb{R}$ be a Lagrangian function and let $c\in \Omega(q_0, q_1)$ and $z_0 \in \mathbb{R}$. Then, $(c,\dot{c}, \mathcal{Z}(c))$ satisfies the Herglotz equations:
    \begin{equation*}  
            \frac{\dd}{\dd t} \frac{\partial L}{\partial \dot{q}^i} 
            - \frac{\partial L}{\partial q^i}=
            \frac{\partial L}{\partial \dot{q}^i} \frac{\partial L}{\partial z},
    \end{equation*}
    if and only if $c$ is a critical point of $\mathcal{A}$.
\end{theorem}

% \
\subsection{An alternative formulation of Herglotz variational principle}
Another way to approach this problem is to consider a constrained variational principle for curves on $Q \times \mathbb{R}$ constrained to a hypersurface $N$. We see that this is equivalent to the dynamics produced by a Lagrangian $L$ when considering unconstrained curves on $Q$.

We will work on the manifold $\bar\Omega(q_0, q_1,z_0)$  of curves $\bar c = (c, c_z):[0,1]\to Q \times \mathbb{R}$ such that $c(0)=q_0$, $c(1)=q_1$, $c_z(0)=z_0$. We do not constraint $c_z(1)$. The tangent space at the curve $c$ is given by
\begin{equation}
    \begin{split}
        T_c &\bar\Omega(q_0,q_1,z_0) =  \{
            \delta \bar c = (\delta c, \delta c_z):[0,1] \to T(Q \times \mathbb{R}) \mid \\ & \delta \bar  c(t) \in  T_{c(t)} (Q \times \mathbb{R}) \, \text{for all t} \in [0,1],   \delta c(0)=0, \, \delta c(1)=0, \delta c_z(0) = 0  \}.
    \end{split}
\end{equation}
In this space, the action functional $\bar{\mathcal{A}}$ can be defined as an integral
\begin{equation}
    \begin{aligned}
        \bar{\mathcal{A}}: \bar\Omega(q_0, q_1,z_0) &\to \mathbb{R},\\
        \bar{c} & \mapsto z_1 - z_0 = \int_0^1 \dot{c}_z(t)  \dd t.
    \end{aligned}
\end{equation}

We will restrict this action to the set of paths that satisfy $\dot{c}_z = L$. For this, consider the hypersurface $N \subseteq T(Q \times \mathbb{R})$, which is the zero set of the constraint function $\phi$:
\begin{equation}\label{eq:lag_constraints}
    \phi(q,\dot{q},z,\dot{z}) = \dot{z} - L(q,\dot{q},z).
\end{equation}
Conversely, given any hypersurface $N$ transverse to the $\dot{z}$-parametric curves, by the implicit function theorem there exists locally a function $L$ such that $N$ is given by the equation $\dot{z}=L$. In this sense, we see that an hypersurface $N\subseteq T(Q \times \mathbb{R})$ is roughly equivalent to a Lagrangian $L:TQ \times \mathbb{R} \to \mathbb{R}$.

We consider the submanifold of curves tangent to $N$
\begin{equation}
    \bar\Omega_N (q_0, q_1,z_0)= \{\bar{c} \in \bar\Omega(q_0, q_1,z_0) \mid \dot{\bar{c}}(t) \in N \, \text{for all t} \}
\end{equation}

Notice that the map $\Id \times \mathcal{Z}: \Omega(q_0,q_1) \to \bar\Omega_N(q_0,q_1,z_0)$ given by ${(\Id \times \mathcal{Z})}(c) = (c, \mathcal{Z}(c))$ is a bijection, with inverse $\pr_Q (c,c_z) = c$. Here, $\mathcal{Z}$, is defined on~\eqref{eq:Z_operator}. Moreover, the following diagram commutes
\begin{equation}
    \begin{tikzcd}
        & \mathbb{R} &                                                             \\
{\Omega(q_0, q_1)} \arrow[rr, "\Id \times \mathcal{Z}"] \arrow[ru, "\mathcal{A}"] &            & {\bar\Omega_N (q_0, q_1,z_0)} \arrow[lu, "\bar{\mathcal{A}}"']
\end{tikzcd}\end{equation}
Hence $\bar{c} \in \bar\Omega_N (q_0, q_1,z_0)$ is a critical point of $\bar{\mathcal{A}}$ if and only if $c$ is a critical point of $\mathcal{A}$. So the critical points of $\mathcal{A}$ restricted to $\bar\Omega(q_0,q_1,z_0)$ are precisely the curves that satisfy the Herglotz equations.

We will also provide an alternate proof. We find directly the critical points of $\bar{\mathcal{A}}$ restricted to $\bar\Omega_N (q_0, q_1,z_0) \subseteq \bar\Omega (q_0, q_1,z_0)$ using the following infinite-dimensional version of the Lagrange multiplier theorem~\cite[3.5.29]{Abraham1988}.
\begin{theorem}[Lagrange multiplier Theorem]\label{thm:Lagrange_multipliers}
    Let $M$ be a smooth manifold and let $E$ be a Banach space such that $g:M \to E$ is a smooth submersion, so that $A=g^{-1}(\{0\})$ is a smooth submanifold. Let $f:M \to \mathbb{R}$ be a smooth function. Then $p \in A$ is a critical point of $f \vert_A$ if and only if there exists $\hat{\lambda} \in E^*$ such that $p$ is a critical point of $f + \hat{\lambda} \circ g$. 
\end{theorem}

We will apply this result to our situation. In the notation of this last theorem, $M= \bar{\Omega}(q_0,q_1,z_0)$ is the smooth manifold. We pick the Banach space $E = L^2([0,1] \to \mathbb{R})$ of square integrable functions. This space is, indeed, a Hilbert space with inner product
\begin{equation}
    \langle \alpha, \beta \rangle = \int_0^1 \alpha(t) \beta (t) \dd t.
\end{equation}
We remind that, by the Riesz representation theorem, there is a bijection between $L^2([0,1] \to \mathbb{R})$ and its dual such that for each $\hat{\alpha} \in L^2([0,1] \to \mathbb{R})^*$ there exists $\alpha \in L^2([0,1] \to \mathbb{R})$ with $\hat{\alpha}(\beta)=  \langle \alpha, \beta \rangle$ for all $\beta \in L^2([0,1] \to \mathbb{R})$.

Our constraint function is
\begin{equation}
    \begin{aligned}
        g: \bar{\Omega}(q_0,q_1,z_0) &\to  L^2([0,1] \to \mathbb{R}),\\
        \bar{c} &\mapsto
         (\phi) \circ (\bar{c}, \dot{\bar{c}}),
    \end{aligned}
\end{equation}
where $\phi$ is a constraint locally defining $N$.
Note that $A=g^{-1}(0) = \bar{\Omega}_N(q_0,q_1, z_0)$.

By Theorem~\ref{thm:Lagrange_multipliers}, $c$ is a critical point of $f=\bar{\mathcal{A}}$ restricted to $\bar{\Omega}_N (q_0,q_1,z_0)$ if and only if there exists $\hat{\lambda} \in L^2([0,1] \to \mathbb{R})^*$ (which is represented by $\lambda \in L^2([0,1] \to \mathbb{R})$) such that $c$ is a critical point of $\bar{\mathcal{A}}_\lambda = \bar{\mathcal{A}} + \hat{\lambda} \circ g$.

Indeed,
\begin{equation}
    \bar{\mathcal{A}}_\lambda = \int_0^1 \mathcal{L}_\lambda(\bar{c}(t), \dot{\bar{c}}(t)) \dd t,
\end{equation}
where
\begin{equation}
    \mathcal{L}_\lambda(q,z,\dot{q},\dot{z})= \dot{z} - \lambda \phi(q,z, \dot{q},\dot{z}).
\end{equation}

Since the endpoint of $c_z$ is not fixed, the critical points of this functional $\bar{\mathcal{A}}_\lambda$ are the solutions of the Euler-Lagrange equations for $\mathcal{L}_\lambda$ that satisfy the natural boundary condition:
\begin{equation}
    \frac{\partial \mathcal{L}_\lambda}{\partial \dot{z}}(\bar{c}(1),\dot{\bar{c}}(1)) = 1- \lambda(1) \frac{\partial \phi}{\partial \dot{z}}(\bar{c}(1),\dot{\bar{c}}(1)) = 0.
\end{equation}
For $\phi = \dot{z}-L$, this condition reduces to $\lambda(1)=1$.

The Euler-Lagrange equations of $\mathcal{L_\lambda}$ are given by
\begin{subequations}\label{eq:euler_langrange_contact}
    \begin{align} \label{eq:euler_langrange_q}
        \frac{\dd }{\dd t} \left(\lambda(t)  
        \frac{\partial \phi(\bar{c}(t),\dot{\bar{c}}(t))}{\partial \dot{q}^i}  \right) - \lambda(t) \frac{\partial \phi(\bar{c}(t),\dot{\bar{c}}(t))}{\partial q^i} &= 0 \\
        \frac{\dd }{\dd t} \left(\lambda(t)     \label{eq:euler_langrange_z}
        \frac{\partial  \phi(\bar{c}(t),\dot{\bar{c}}(t))}{\partial \dot{z}}  \right) - \lambda(t) \frac{\partial \phi(\bar{c}(t),\dot{\bar{c}}(t))}{\partial z} &= 0,
    \end{align}
\end{subequations}
since $\phi = \dot{z} - L$, the equation~\eqref{eq:euler_langrange_z} for $z$ is just
\begin{equation}
    \frac{\dd \lambda(t)}{\dd t} = - \lambda(t) \frac{\partial L}{\partial z},
\end{equation}
substituting on \eqref{eq:euler_langrange_q} and dividing by $\lambda$, we obtain Herglotz equations.

\subsection{Vakonomic constraints}
If we have more constraints, we can obtain vakonomic dynamics, just by changing $\phi$ by $\phi^a$ and $\lambda$ by $\lambda_a$ on \eqref{eq:euler_langrange_contact}, where $a$ ranges from $0$ to the number of constraints $k$. Indeed, we restrict our path space to the ones tangent to submanifold $\tilde{N} \subseteq N \subseteq T(Q \times \mathbb{R})$, where $N$ is the zero set of $\phi^0$, given by $\phi^0 = \dot{z} - L$. Repeating the similar computations, we would find that the critical points of $\mathcal{A}\vert_{\Omega(q_0,q_1,\tilde{N})}$ are the solutions of
    \begin{subequations} \label{eq:euler_langrange_contact_vakonomic_2}
    \begin{align} \label{eq:euler_langrange_vakonomic_q}
        \frac{\dd }{\dd t} \left(\lambda_a(t)  
        \frac{\partial \phi^a(\bar{c}(t),\dot{\bar{c}}(t))}{\partial \dot{q}^i}  \right) - \lambda_a(t) \frac{\partial \phi^a(\bar{c}(t),\dot{\bar{c}}(t))}{\partial q^i} &= 0 \\
        \frac{\dd }{\dd t} \left(\lambda_a(t)     \label{eq:euler_langrange_vakonomic_z}
        \frac{\partial  \phi^a(\bar{c}(t),\dot{\bar{c}}(t))}{\partial \dot{z}}  \right) - \lambda_a(t) \frac{\partial \phi^a(\bar{c}(t),\dot{\bar{c}}(t))}{\partial z} &= 0,\\
        \phi^a(\bar{c}(t),\dot{\bar{c}}(t))&=0,
        \end{align}
    \end{subequations}
    where $(\phi^a)_{a=0}^k$ are constraints defining $\tilde{N}$ as a submanifold of $TQ \times \mathbb{R}$. Since $\frac{\partial \phi^0}{\partial \dot{z}} = 0$, the rest of the constraints can be chosen to be independent of $\dot{z}$. We denote
    \begin{align}
        \psi^\alpha(q,\dot{q},z) &= \phi^\alpha(q,\dot{q},z,L(q,\dot{q},z)),\\
        \mu_\alpha &= \frac{\lambda_\alpha}{\lambda_0}\\
        \mathcal{L}_\mu (q,\dot{q},z,t) &= L(q,\dot{q},z) - \mu_\alpha(t) \psi^\alpha(q,\dot{q},z) 
    \end{align}
    for $\alpha \in \{1, \ldots k \}$, provided that $\lambda_0 \neq 0$.

From this, we can write the equations~\eqref{eq:euler_langrange_contact_vakonomic_2} as
\begin{subequations} \label{eq:euler_langrange_contact_vakonomic_3}
    \begin{align} \label{eq:euler_langrange_vakonomic_q_3}
        -\frac{\dd }{\dd t} \left(\lambda_0(t) \frac{\partial \mathcal{L}_\mu }{\partial \dot{q}^i}  \right) 
        + \lambda_0(t) \frac{\partial \mathcal{L}_\mu }{\partial q^i} &= 0 \\ \label{eq:euler_langrange_vakonomic_z_3}
        \frac{\dd \lambda_0(t) }{\dd t} = 
         \lambda_0(t) \frac{\partial \mathcal{L}_\mu }{\partial z} &= 0,\\
        \psi^\alpha(\bar{c}(t),\dot{c}(t))&=0,\\
        \dot{c_z}(t) &= \mathcal{L}_\mu (\bar{c}(t), \dot{c}(t),t).
        \end{align}
    \end{subequations}
    Substituting~\eqref{eq:euler_langrange_vakonomic_z_3} onto \eqref{eq:euler_langrange_vakonomic_q_3}, dividing by $\lambda_0$ and reordering terms, we obtain
\begin{subequations}
    \begin{align} \label{eq:herglotz_vakonomic_q}
        \frac{\dd }{\dd t} \left(\frac{\partial \mathcal{L}_\mu }{\partial \dot{q}^i}  \right) 
        - \frac{\partial \mathcal{L}_\mu }{\partial q^i} &=
        \frac{\partial \mathcal{L}_\mu }{\partial \dot{q}^i} \frac{\partial \mathcal{L}_\mu }{\partial z}\\
    \psi^\alpha(\bar{c}(t),\dot{c}(t))&=0,\\
    \dot{c_z}(t) &= \mathcal{L}_\mu (\bar{c}(t), \dot{c}(t),t).
\end{align}
\end{subequations}

We remark that these equations are just the Herglotz equations for the \emph{extended Lagrangian} $\mathcal{L}$:
\begin{equation}
    \begin{aligned}
        \mathcal{L} : T(Q \times \mathbb{R}^k) \times\mathbb{R} &\to \mathbb{R} \\
        (q,\mu,\dot{q},\dot{\mu}, z) &\mapsto L(q,\dot{q},z) - \mu_\alpha \psi^\alpha (q, \dot{q}, z)
    \end{aligned}
\end{equation}

\subsection{Applications to control}
The Herglotz optimal control problem~\cite{deLeon2020a} can be formulated by working on the control bundle $W \times \mathbb{R} \to Q \times \mathbb{R}$, with local coordinates $(x^i,u^a,z)$: the \emph{variables} $x^i$, the \emph{controls} $u^a$ and the \emph{action} $z$. 

The problem consists on finding the curves $\gamma:I=[a,b]\to W$, $\gamma=(\gamma_Q,\gamma_U, \gamma_z)$, such that
\begin{enumerate}
  \item[1)] end points conditions: $\gamma_Q(a)=x_a, \gamma_Q(b)=x_b, \gamma_z(a)=z_0$,
  \item[2)] $\gamma_Q$ is an integral curve of $X$: $\dot{\gamma}_Q=X\circ(\gamma_Q,\gamma_U)$\, , 
  \item[3)] $\gamma_z$ satifies the differential equation $\dot z=F(x,u,z)$, and  
  \item[4)] maximal condition: $\gamma_z(b)$ is maximum over all curves satisfying 1)--3). 
\end{enumerate}

We remark that this can be interpreted as a vakonomic Herglotz principle on $TW$, with constraints given by the control equations $\phi^i =  X^i(x,u,z) - \dot{q}^i$ and the Lagrangian being the cost function $L(x^i,u^a,\dot{q}^i,\dot{u}^a,z)= F(x,u,z)$.  The equations of motion obtained through the contact vakonomic principle coincides with the ones obtained indirectly through Pontryaguin maximum principle in~\cite[Eq.~28]{deLeon2020a}
\begin{subequations}
    \begin{align}
        \dot{q}^i &= X^i,\\
        \dot{\mu}_i &=  \frac{\partial F}{\partial x^i}  - \mu_j \frac{\partial X^j}{\partial x^i}  - \mu_j \left( \frac{\partial F}{\partial z} - \mu_i \frac{\partial X^j}{\partial z}  \right) \\ &= \nonumber
\mu_i \frac{\partial F}{\partial z} - \mu_j \frac{\partial X^j}{\partial x^i} + \frac{\partial F}{\partial x^i} - \frac{\partial X^j}{\partial z} \mu_i \mu_j, \\
        \dot{z} &= F
    \end{align}
   subjected to the constraints
    \begin{equation}
         \frac{\partial F}{\partial u^a} - \mu_j \frac{\partial X^j}{\partial u^a} = 0.
    \end{equation}
\end{subequations}

% \subsection{Other comments}
% \begin{itemize}
%     \item The variations do not fix the endpoints of $z$. Why can we use the Euler-Lagrange equations?. Apparently, the constrain produced by $\dot{q}=L$ fixes this. I need to explain this step better.
%     \item The equations are symmetric. We could have solved for $\lambda$ on the equation of some other $q^k$ and obtain Herglotz equation but interchanging the role of $z$ and $q^k$. Does this has some interpretation? Do we need to distinguish the $z$-coordinate? Can we work on a manifold $\bar{Q} \neq Q \times \mathbb{R}$?
% \end{itemize}

\section*{Acknowledgements}

\hspace{5mm} M. de Le\'on and M. Lainz acknowledge the partial finantial support from MINECO Grants MTM2016- 76-072-P and the ICMAT Severo Ochoa project SEV-2015-0554.  
M. Lainz wishes to thank MICINN and ICMAT for a FPI-Severo Ochoa predoctoral contract PRE2018-083203.
M.C. Mu\~noz-Lecanda acknowledges the financial support from the 
Spanish Ministerio de Ciencia, Innovaci\'on y Universidades project
PGC2018-098265-B-C33
and the Secretary of University and Research of the Ministry of Business and Knowledge of
the Catalan Government project
2017--SGR--932.

\bibliographystyle{amsplain}
\bibliography{main}
\end{document}